\documentclass[preprint, pra, amsmath, aps,showpacs]{revtex4}
\usepackage{color}
\newcommand{\bea}{\begin{eqnarray} }
\newcommand{\eea}{\end{eqnarray}}
\newcommand{\bean}{\begin{eqnarray*}}
\newcommand{\eean}{\end{eqnarray*}}
\newcommand{\nn}{\nonumber \\}
\def\bfg#1{{\mbox{\boldmath $#1$}}}
\def\A{{\bf A}}
\def\B{{\bf B}}
\def\b{{\bf b}}

\def\E{{\bf E}}

\def\g {{\bf g}}

\def\v{{\bf v}}

\def\X {{\bf X}}

\def\btimes{~{\bf \times}~}
\def\bnabla{{\bf \nabla}}
\def\bcdot{~{\bf \cdot}~}
\newcommand{\lbs}{\left (}
\newcommand{\rbs}{\right )}
\newcommand{\lbm}{\left\lbrack}
\newcommand{\rbm}{\right\rbrack}

\def\od#1,#2{\frac{d#1}{d#2}}
\def\odz#1,#2{\frac{d^2#1}{d{#2}^2}}
\def\pd#1,#2{\frac{\partial #1}{\partial #2}}
\def\pdz#1,#2{\frac{\partial^2 #1}{\partial {#2}^2}}

\def\pdd#1,#2{\frac{\partial^3 #1}{\partial {#2}^3}}
\def\pdv#1,#2{\frac{\partial^4 #1}{\partial {#2}^4}}
\def\pdzz#1,#2,#3{\frac{\partial^2 #1}{\partial {#2}\partial{#3}}}

\def\eq#1{Eq.~(\ref{#1})}
\def\eqn#1{(\ref{#1})}
\usepackage{graphicx}
\usepackage{dcolumn}
\usepackage{bm}

\begin{document}

\hfill{To be published in Physics of Plasmas 2023}

\bibliographystyle{unsrt}
%
%
\title{Modification of Lie's transform perturbation theory \\for 
charged particle motion in a magnetic field
}
%
%
\author{Linjin  Zheng,\email{lzheng@austin.utexas.edu}}
\affiliation{Institute for Fusion Studies,
University of Texas at Austin,
Austin, TX 78712}

\begin{abstract}

It is pointed out that the conventional  Lie transform perturbation theory
for the guiding center motion of charged particles in a magnetic field needs to be
modified for ordering inconsistency. There are two reasons. First,
the ordering difference between  the temporal variation of gyrophase and that of the other 
phase space coordinates needs to be taken into account. Second, 
it is also important to note that  
the parametric limit of the derivative of a function is not equivalent to  the derivative of the limit
function.  When these facts are taken into account,
 the  near identity transformation rule for one form related to the Lagrangian is modified.
With the modified near identity transformation rule, the 
drift motion of  charged particles can be described in the first
order, instead of in the second order and beyond through a tedious expansion process as in the conventional formulation.    
This resolves the discrepancy between the direct and Lie transform treatments 
in the Lagrangian perturbation theory for charged particle motion in a magnetic field.

\end{abstract}

\pacs{52.53.Py, 52.55.Fa, 52.55.Hc}

\maketitle

 \section{Introduction}

 The guiding center motion of charged particles in a magnetic field is a fundamental topic 
 for magnetically confined fusion research. The standard guiding center theory has been developed 
 since  the 1950s  as reviewed in reference \cite{northrop63}.
The topic was later revisited using the Hamiltonian and Lagrangian theories,
for example in references \cite{boozer,white,littlejohn83,cary83,cary09}.  
This is mainly related to the development
of nonlinear gyrokinetic simulation as studied or reviewed in
 references \cite{hahm,qin,brizard07}. This is due to the concerns that
the  standard guiding-center theory 
does not preserve Liouville’s theorem \cite{gold} and
the energy conservation law for time-independent systems.
 These properties are especially important for long-time simulations. 
This has motived the further development of the Hamiltonian and Lagrangian theories for 
guiding center motion.

One of  the important developments in the  Hamiltonian and Lagrangian theories
for guiding center motion of charged particles is the introduction of the
phase space  Lagrangian theories and the Lie transform perturbation technique.
The pioneer contributions in this procedure can be found in references \cite{littlejohn83,cary83,cary09,iib.arnold89}
and references therein. 
The detailed Lie transformation procedure for guiding center motion of charged particles, 
which were omitted in the original work in reference \cite{littlejohn83},
were given in reference \cite{brizard98}. The Lie transform method
 provides a systematic perturbation theory for guiding center motion
 and has many advantages, such as 
the near identity transform process allowing
the expansion generator  be determined in the order-by-order analyses. 
Also, the backward transformation can be obtained easily from the forward transformation.

However, there is a discrepancy between the direct and Lie transform treatments 
in the Lagrangian perturbation theory in  phase space for charged particle motion in a magnetic field. In the direct method, the phase space Lagrangian valid to
the first order is given as follows  \cite{cary09}
\bea
{^d} \Gamma = \lbs   \frac{e}{mc}\A +u\b\rbs \bcdot d\X
+  \frac{mc}e\mu d \zeta - \lbs  \frac {u^2}2+\mu B + \frac{e}m \varphi\rbs  dt,
\label{iib8.l4}
\eea
while the standard Lie transform theory \cite{littlejohn83}, which is  detailed in the appendix
of reference \cite{brizard98}    (Eq.~(B18) with the zeroth order contribution added), yields in the same order
\bea
{^d}\Gamma &=&  \lbs    \frac{e}{mc} \A +u\b\rbs \bcdot d\X - \lbs  \frac {u^2}2+\mu B +
 \frac{e}m \varphi\rbs  dt.
\label{iib8.bgamma1}
\eea 
Here, 
the general phase space
coordinate system $\vec Z =\{\X, \mu,   u,\zeta;t\} $ is used,  $\X$ is related to the guiding center coordinate,
$\mu = v_\bot^2/2B$ is the magnetic moment,   $\v$ denotes the  particle velocity
with $v_\bot$ being the perpendicular component and 
$u$  the parallel component,  and
$t$ represents time, $\A$ is the vector potential,  $\B$ is the magnetic field, $\b=\B/B$, 
  $\varphi$ denotes the electric scalar potential,  $e$ is the charge, $c$ is the speed of light,
the boldface denotes the vector   in configuration space, $\bar {(\cdot)}$ and $\vec {(\cdot)}$ are introduced to represent respectively  the covariant and contravariant vectors in the phase space with time included.   
The constant factor $m$ for Lagrangian, for example  in
Eqs.  \eqn{iib8.l4} and \eqn{iib8.bgamma1}, has been cast aside.   To distinguish one form   
(a covariant vector)$\bar\Gamma =\{\Gamma_\mu\} $ and $\Gamma_\mu dZ^\mu$, the notation for  the scalar zero form  $^d\Gamma = \Gamma_\mu dZ^\mu$ is introduced.  In these analyses, the gyrofrequency
$\Omega=eB/mc$ is assumed to be high, i.e., $(v/R) /\Omega \sim (\partial/\partial t) /\Omega
\sim {\cal O} (\epsilon) \ll 1$, where
$R$ is the scale of electromagnetic field, which is  larger than the Larmor radius by an order of magnitude. 
It is also assumed that $v_\E =|\E\times\B|/B^2 \ll v$.

Comparing Eqs. \eqn{iib8.l4} and \eqn{iib8.bgamma1}
 one can see that the term ``$  (mc/e)\mu d \zeta$" is missing in the   conventional Lie transformation treatment in the first order.
   Note that
\bea
\frac{ (mc/e)\mu d \zeta}{u\b\bcdot d\X}&\sim& \frac{ (mc/e)(v_\bot^2/B) (d \zeta/dt)}{u\b\bcdot (d\X/dt)}
\nn
&\sim& \frac{(v_\bot^2/(eB/m  c)) (d \zeta/dt)}{u\b\bcdot (d\X/dt)}\sim\frac {v_\bot^2}{  u^2}\sim 1.
\label{p3}
\eea
Here, it has been used that $d \zeta/dt \sim eB/m  c$ and $d\X/dt\sim \v$.
The ordering estimate in \eq{p3} shows that one cannot regard $ (mc/e)\mu d \zeta$ as $O(\epsilon)$ as compared to $u\b\bcdot d\X$.
 Equation \eqn{iib8.bgamma1} is therefore ordering inconsistent because the term $u\b \bcdot d\X$ is kept  but the term  $  (mc/e)\mu d\zeta$ is dropped,
  noting that they are of the same order.
  
      Note that  the detailed derivation process
  was omitted   in  Ref. \cite{littlejohn83} . This leads us to discuss directly Ref.  \cite{brizard98}.
Nevertheless, as pointed in Ref.  \cite{brizard98}, the detailed derivation given in the Appendix B
of the paper is similar to that in  Ref. \cite{littlejohn83} except the rotation effects being added.
It is especially noted  that the ordering inconsistency in 
  Eq.~(B18) of  Ref. \cite{brizard98} discussed above appears also  in Eq.~(20) of Ref. \cite{littlejohn83}.
Like a mathematical theorem, it persists unless a counter-proof is given. 
The Lie transform theory has become  a standard perturbation theory after  thorough reviews for example 
in  the journal Reviews of modern physics
(see for instance Refs. \cite{cary83,brizard07,cary09,iib.arnold89}).
From the citation list of relevant articles one can  see that the classical works in Lie transform
are still actively used. 

As will be seen, this is because 
in  the conventional Lie transform perturbation theory the following type of deduction has been employed 
\bea
\left. \pd   F,Z\right|_{\epsilon=0} =  \pd {\left.   F\right|_{\epsilon=0}} ,Z.
\label{eq4}
\eea
Apparently, this is not generally applicable. The parametric limit of the derivative of a function is not equivalent to  the derivative of the limit
function.  This becomes serious for the system with rapidly varying coordinate. 
For example
for the case with  $   F=   \left. F\right|_{\epsilon=0} +\epsilon    f (Z)$ and $\partial    f /\partial Z\sim 1/\epsilon$, 
one has  
\bea
\left. \pd   F,Z\right|_{\epsilon=0} =  \pd {\left.    F \right|_{\epsilon=0}} ,Z + \pd    f,Z\not  
=  \pd {\left.    F\right|_{\epsilon=0}} ,Z.
\eea
The terms $\pd {\left.    F\right|_{\epsilon=0}} ,Z$ and $\pd    f,Z$ are actually of the same order in this case.
The Larmor radius expansion exactly has this feature. Although the Larmor radius is small, but
 the gyrophase 
varies   in time rapidly.
 
It is interesting to point out that a similar case in gyro-fluid models was  pointed out earlier in Ref. \cite{mac}
and further clarified later in Refs. \cite{mik,del}.
It is also related to the use of the
parametric expansion with respect to the small parameter  $(\partial /\partial t) /\Omega\ll1$. 
When it is applied to the second order fluid moment of Vlasov equation the inclusion of first order Finite-Larmor-Radius (FLR) corrections to a double-adiabatic closure in a fluid modelling causes the dispersion
relation of magneto-acoustic waves propagating perpendicularly to a background
magnetic field to get a wrong sign in their spatial dispersion.
The unphysical result is fixed only if the $\epsilon\ll1$ expansion is directly performed on the fluid moment equations which account for the full anisotropy of the second and third order velocity moments.

In this paper, we modify the conventional Lie transformation formalism 
for the guiding center motion of charged particles in a magnetic field
by taking into account  the ordering difference between  the temporal variations of gyrophase and that of the other 
phase space coordinates and the fact that
the parametric limit of the derivative of a function is not equivalent to  the derivative of the limit function. 
This leads us to resolve the discrepancy between the direct and Lie transform treatments 
in the Lagrangian perturbation theory in the phase space. The modification presented in this paper is expected to affect generally
the  Lie transform perturbation formulation for the systems with a rapidly varying coordinate. 

The manuscript is organized as follows. In Sec. II a brief review of the conventional Lie transform theory
is given; in Sec. III. the modification of Lie's transform perturbation theory for 
charged particle motion in a magnetic field is described; In the last section, conclusions
and discussion are presented. Appendix A is introduced to double confirm the newly
derived transformation rule for one form.

 \section{Review of the conventional Lie transform perturbation theory}

In this section, we briefly  review the conventional Lie transform perturbation theory for
the charged particle motion in a magnetic field \cite{littlejohn83,cary09}. This
will pave the way for the modified theory to be described in the next section.

With a small parameter $\epsilon$, the near-identity coordinate transform can be generally expressed as
\bean
Z^\mu&=& z^\mu + \epsilon Z_{1f}^\mu(\vec z) + \epsilon^2 Z_{2f}^\mu(\vec z)+ \cdots
\eean
Here, the subscript ``f" denotes the forward transformation from the current
 to the new coordinates.  For
the charged particle motion in a magnetic field, the forward transformation
is simply the transformation to the guiding center. 
 The forward transformation can be generally denoted as 
\bea
Z^\mu=Z_f^\mu(\vec z,\epsilon).
\label{ii7.forw}
\eea
 We also introduce   
the backward transformation as 
\bea
z^\mu=Z_b^\mu(\vec Z_f(\vec z,\epsilon),\epsilon).
\label{ii7.back}
\eea

In the Lie transformation, the coordinate transformation is specified through a generator $g^\mu$ such that
\bea
\pd {Z_f^\mu(\vec z,\epsilon)},\epsilon &=& g^\mu(\vec Z_f(\vec z,\epsilon)) ~~~\hbox{and}~~~ Z_f^\mu(\vec z,0) =z^\mu.
\label{ii7.fde}
\eea 
Applying $\partial /\partial \epsilon$ on the backward transformation in \eq{ii7.back}, one obtains
\bean
\pd {Z_b^\mu},{Z_f^\nu} \pd {Z_f^\nu},\epsilon + \pd {Z_b^\nu},\epsilon =0.
\eean
Here, the summation for repeated indices is  implied as usual.
Using \eq{ii7.fde}, one obtains
\bea
\pd {Z_b^\mu(\vec z,\epsilon)},\epsilon &=& - g^\nu(\vec Z) \pd {Z_b^\mu},{Z_f^\nu}.
\label{ii7.bde}
\eea

We first consider  the application of the Lie transformation on a scalar. Suppose there is a forward transform from a scalar $s(\vec z)$ to 
a new scalar $S(\vec Z,\epsilon)$ such that
\bea
S(\vec Z,\epsilon) = s(\vec Z_b(\vec Z,\epsilon)).
\label{ii7.scalar}
\eea
Applying $\partial /\partial \epsilon$ on it, one obtains
\bean
\pd {S(\vec Z,\epsilon)},\epsilon  = \pd {s(\vec Z_b(\vec Z,\epsilon))},{Z^\mu_b} \pd{Z^\mu_b},\epsilon
=\pd {S(\vec Z,\epsilon)},{Z^\mu_b} \pd{Z^\mu_b},\epsilon.
 \eean
Using \eq{ii7.bde} and the chain rule, one obtains
\bea
\pd {S(\vec Z,\epsilon)},\epsilon  =  - g^\mu(\vec Z) \pd {S(\vec Z,\epsilon)},{Z^\mu} .
\label{ii7.dgd}
 \eea
Defining the operator $L_g\equiv g^\mu(\partial/\partial Z^\mu)$. \eq{ii7.dgd} becomes
\bea
\pd S,\epsilon  = -L_g S.
\label{ii7.sde}
 \eea
This further gives that
\bea
\pd^n S,{\epsilon^n}  = \lbs -L_g \rbs^n S.
\label{ii7.sden}
 \eea

We now expand $S(\vec Z,\epsilon)$ in a Taylor series:
\bean
S(\vec Z,\epsilon) &=&\left. \sum_{n=0}^{+\infty}\frac{\epsilon^n}{n!} \pd^n S,{\epsilon^n}\right |_{\epsilon=0}.
\eean
  If the limit   and derivative are  commutable, this equation becomes
\bea
S(\vec Z,\epsilon) &=& \sum_{n=0}^{+\infty}\frac{\epsilon^n}{n!} \pd^n S  |_{\epsilon=0},{\epsilon^n}
= \sum_{n=0}^{+\infty}\frac{\epsilon^n}{n!} \pd^n s ,{\epsilon^n}.
\label{ii7.tay}
\eea
Using Eqs. \eqn{ii7.sden} and   \eqn{ii7.tay},
one obtains the Lie transformation of a scalar:
\bea
S= e^{ -\epsilon L_g } s.
\label{ii7.sct}
 \eea
As pointed out in the introduction section  that
the parametric limit of the derivative of a function is not equivalent to  the derivative of the limit
function, this transformation rule needs to be verified case by case.
Because of the operator expression, the inverse transform is simply
\bea
s= e^{ \epsilon L_g } S,
\eea
evaluating at $\epsilon =0$.

The coordinate transformation defined in equations \eqn{ii7.forw} and \eqn{ii7.back} 
indicates that $z^\alpha= Z_b^\alpha(\vec Z,\epsilon)$. Comparing with \eq{ii7.scalar},
one can see that the coordinate transformation  is just a special case of scalar transformation. 
Let us introduce the coordinate function $I^\alpha$ defined by $I^\alpha(\vec z) = z^\alpha$
for a particular $\alpha$.  One then has
\bea
Z_b^\alpha = e^{-\epsilon L_g } I^\alpha(\vec z).
\label{zzz0}
\eea
Explicitly, one has   \cite{brizard98}
\bea
  Z^\alpha(\vec Z,\epsilon)&=&   z^\alpha
-\epsilon L_g z^\alpha +\frac12 \epsilon^2 L_g  \lbs  L_g z^\alpha \rbs +\cdots
\nn
&=&  z^\alpha-\epsilon g_1^\alpha-\epsilon^2\lbs
g_2^\alpha-\frac12 g_1^\beta\pd g_1^\alpha,{  z^\beta}\rbs +\cdots,
\label{ii7.gtol}
\eea
where $g_{\cdots}$ are the functions of $  \vec z$.
 Since the transformation is invertible, the inverse transformation
is given by
\bea
  z^\alpha(\vec z,\epsilon)&=&  Z^\alpha+\epsilon g_1^\alpha+\epsilon^2\lbs
g_2^\alpha+\frac12 g_1^\beta\pd g_1^\alpha,{ Z^\beta}\rbs +\cdots,
\label{ii7.ltog}
\eea
where $g_{\cdots}$ are the functions of $ \vec Z$.

Next, we  consider the Lie transform on a 1-form in the phase space: $\gamma_\mu$.
For a coordinate transformation $\vec z\to \vec Z$, the new 1-form $\Gamma_\mu$
follows the invariant  relation: $\Gamma_\mu dZ^\mu=\gamma_\mu dz^\mu$. This is completely
equivalent to the usual rule for transforming a covariant vector:
\bea
\Gamma_\mu (\vec Z,\epsilon) &=&  \pd {Z_b^\nu(\vec Z,\epsilon)} ,{Z^\mu}\gamma_\nu(\vec Z_b(\vec Z,\epsilon)).
\label{bigsmall}
\eea
Applying $\partial /\partial \epsilon$ on it and using \eq{ii7.dgd}, one obtains
\bea
\pd {\Gamma_\mu (\vec Z,\epsilon)},\epsilon &=& - \pd ,{Z^\mu} \lbm g^\lambda(\vec Z)\pd {Z_b^\nu(\vec Z,\epsilon)},{Z^\lambda}   \rbm \gamma_\nu(\vec Z_b(\vec Z,\epsilon))
\nn&&
-g^\lambda(\vec Z)  \pd {Z_b^\nu(\vec Z,\epsilon)} ,{Z^\mu}
\pd {\gamma_\nu(\vec Z_b(\vec Z,\epsilon))},{Z^\lambda}   
\nn
&=&
- g^\lambda(\vec Z) \lbm      \pd {\Gamma_\mu(\vec Z,\epsilon)} ,{Z^\lambda}
-   \pd {\Gamma_\lambda(\vec Z,\epsilon)} ,{Z^\mu}
\rbm -   \pd  ,{Z^\mu} \lbm  g^\lambda(\vec Z) \Gamma_\gamma(\vec Z,\epsilon) \rbm .
\label{ii7.d1f0}
\eea
Let $\xi_\mu$ be an arbitrary 1-form and $L_g\bar \xi $ another 1-form whose components given by
\bea
(L_g\bar \xi)_\mu &=&  g^\nu\lbs \partial_\nu\xi_\mu    -\partial_\mu\xi_\nu \rbs.
\label{ii7.for}
\eea
  To be specific, we point out that \eq{ii7.for}    is just Eq.(46) in Ref. \cite{cary83}, in which
 the last term of \eq{ii7.d1f0} has been dropped  in view of that  the total derivative does not
contribute to the variational principle.
With this definition, \eq{ii7.d1f0} can be expressed as
\bea
\pd\bar  \Gamma,\epsilon&=&-L_g\bar \Gamma-\bar \partial (\vec g\bcdot\bar \Gamma),
\label{ii7.d1f1}
\eea
  which is just Eq.(47) in Ref. \cite{cary83}. Here, $\bar\partial = \{\partial /\partial Z^\mu\}$.

Noting the symmetry property, one can prove that $L_g \bar \partial $ always vanishes (similar to $\bnabla\btimes\nabla =0$
in the three dimensional case)
and $g^\mu(L_g\bar \xi)_\mu =0$. Therefore, \eq{ii7.d1f1} can be applied inductively to yield
\bea
\pd^n\bar \Gamma,{\epsilon^n}&=&\lbs -L_g\rbs^n \bar \Gamma+\lbs -\bar \partial \vec g\bcdot\rbs^n \bar \Gamma.
\label{ii7.d1f}
\eea
  If the limit  and and derivative are  commutable, \eq{ii7.d1f} leads to
\bean
\bar \Gamma(\vec Z,\epsilon) &=&\left. \sum_{n=0}^{+\infty}\frac{\epsilon^n}{n!} \pd^n \bar \Gamma ,{\epsilon^n} \right|_{\epsilon=0} = \sum_{n=0}^{+\infty}\frac{\epsilon^n}{n!} \pd^n \bar \Gamma  |_{\epsilon=0},{\epsilon^n}
= \sum_{n=0}^{+\infty}\frac{\epsilon^n}{n!} \pd^n \bar \gamma ,{\epsilon^n}.
\eean
Using this equation and \eq{ii7.sden}, one obtains
\bea
\bar  \Gamma&=&e^{-\epsilon L_g}\bar\gamma +\bar \partial S 
\label{ii7.1ft}
\eea
Here,    $\bar\partial S $  results from 
the second term on the right hand side of \eq{ii7.d1f} and
gives rise to an exact differential in the variational principle.
The inverse of \eq{ii7.1ft} is simply
 \bea
\bar \gamma&=&e^{\epsilon L_g}\bar \Gamma +\bar \partial s, 
\label{ii7.1ftr}
\eea
where $s$ is a scalar different from $S$. As pointed out in the introduction section  that
the parametric limit of the derivative of a function is not equivalent to  the derivative of the limit
function, this transformation rule needs to be verified case by case. 

Using the transformation formula for scalar and 1-form in equations \eqn{ii7.sct}   
and \eqn{ii7.1ft} (together with \eq{ii7.for}) respectively, one can develop the high order
perturbation theory. We assume that the 1-form  in the phase space which can be expanded as follows
\bea
\bar \gamma =\bar  \gamma^{(0)} +\epsilon \bar  \gamma^{(1)}+\epsilon^2 \bar  \gamma^{(2)}+\cdots.
\label{ii7.1fn}
\eea
It is also assumed that the  lowest-order dynamics with $\bar\gamma^{(0)}$ is well understood, i.e.,
its solutions are known or can be easily obtained. Therefore, the lowest-order trajectory can be used 
to find the solutions of higher orders. 

In order to simplify the 1-form to sufficient orders, the following overall transformation operator,
which is a composition of individual Lie transforms,
is introduced
\bea
T&=& \cdots T_3T_2T_1
\eea 
with
\bea
T_n&=& e^{-\epsilon^n L_n}.
\eea
Here, $L_n$ denotes $L_{g_n}$ as defined
for scalar and 1-form in equations \eqn{ii7.sct}   
and \eqn{ii7.1ft} (together with \eq{ii7.for}) respectively. The generators $g^\mu_n$ with
$n=1,\cdots,n$ will be  used to simply the fundamental 1-form in \eq{ii7.1fn} to order $n$.
The inverse of the transformation vector is
\bea
T^{-1}&=&  T_1^{-1}T_2^{-1}T_3^{-1} \cdots
\eea 
with
\bea
T_n^{-1}&=& e^{\epsilon^n L_n}.
\eea

When successive Lie transforms are applied in this manner,   \eq{ii7.1ft} becomes,
noting that $T_i \bar\partial $ always vanishes,  
\bea
\bar \Gamma&=&T\bar \gamma +\bar \partial S ,
\label{ii7.1ftn}
\eea
where $S$  collects all  possible scalar contributions.  Expanding $\bar\Gamma$ and $S$ in powers of
$\epsilon$ as well as $\bar\gamma$ in \eq{ii7.1fn} and collecting terms in each order, one obtains
\bea
\bar \Gamma^{(0)} &=& \bar \gamma^{(0)} ,
\label{lag0th}
\\
\bar \Gamma^{(1)} &=& \bar\partial S^{(1)}  -L_1  \bar \gamma^{(0)} +  \bar \gamma^{(1)},
\label{lag1st}
\\
\bar \Gamma^{(2)} &=&\bar  \partial S^{(2)}  -L_2  \bar \gamma^{(0)} +  \bar \gamma^{(2)} - L_1  \bar\gamma^{(1)}
+\frac12  L_1^2  \bar \gamma^{(0)},
\\
\bar \Gamma^{(3)} &=& \bar \partial S^{(3)}  -L_3  \bar \gamma^{(0)} + \bar \gamma^{(3)} - L_2L_1 \bar  \gamma^{(0)}
+\frac16  L_1^3 \bar  \gamma^{(0)}
\nn&&
- L_2 \bar  \gamma^{(1)}
+\frac12  L_1^2  \bar \gamma^{(1)} -  L_1 \bar  \gamma^{(2)},
\label{lag2nd}
\eea
and so on.

  These complete the basic theoretical review of the conventional Lie transform theory in references \cite{littlejohn83,cary09}.
The conventional Lie transform theory was applied
to study the charged particle motion in a magnetic field \cite{littlejohn83,brizard98}.
In the Appendix B  of Ref. \cite{brizard98},    the Lagrangian in the zeroth order was obtained as follows
\bea
^d\Gamma^{(0)} &=&  \frac{e}{mc}   \A  \bcdot d\X,
\label{zero}
\eea
while the first order Lagrangian is given as follows  (Eq.~(B18) in Ref.  \cite{brizard98})
\bea
^d\Gamma^{(1)} &=&  u\b \bcdot d\X - \lbs  \frac {u^2}2+\mu B +  \frac{e}{m} \varphi\rbs  dt.
\label{first}
\eea
Combining Eqs. \eqn{zero} and \eqn{first} yields \eq{iib8.bgamma1} discussed in the introduction.

  In these reviews, the correction as pointed out in the introduction section has not been included, especially 
 as will be seen in the next section, the expansions of the phase space Lagrangian in
Eqs. \eqn{lag0th} - \eqn{lag2nd} will be modified.   Consequently, the first order Lagrangian in \eq{first} will
be modified.     The modification is related to the difference between \eq{first} and the result
by the direct approach in \eq{iib8.l4} .   The term $  (mc/e)\mu d\zeta$ is missing
in \eq{first}, although it is of the same order as the  term $u\b \bcdot d\X$ as shown 
in the ordering analysis in  \eq{p3}.  The term $  (mc/e)\mu d\zeta$ is only picked
up in the next order in the conventional Lee transform approach  (Eq.~(B30) in Ref.  \cite{brizard98}
  or Eq.(29) in Ref. [4]). 
Note that   \eq{first} is obtained by strictly following the standard Lie transform formulation. As explained
in the next section, the problem lies in that the  commutation between
the limit and derivative in deriving \eq{ii7.1ft} is  illegitimate in the case with a fast varying coordinate.
In the derivation of \eq{ii7.1ft}, the commutation  as shown 
in \eq{eq4} was used to reduce $\bar \Gamma$  to $\bar \gamma$ in the conventional Lie transform formulation. 
As explained alternatively in Appendix \ref{app}, the illegitimate commutation is equivalent to   
assume that $\Gamma_\mu dZ^\mu =\gamma_\mu dZ^\mu$, which is apparently invalid. The correct
one should be $\Gamma_\mu dZ^\mu =\gamma_\mu dz^\mu$.

 \section{Modification of Lie transformation}

In this section, we describe the modification of the conventional Lie transformation formulation 
 for the systems containing rapidly varying coordinates. 
For the guiding center motion of a charged particle in a magnetic field, the gyrophase is 
this type of  coordinates.   This helps solve the inconsistency in the phase-space Lagrangian
perturbation theories between the direct derivation and the Lie transform formulation
as pointed out in the introduction section. 

Strictly speaking, the transformation rule for 1-form in \eq{ii7.for} is correct. 
However, when applying this rule in  equations \eqn{lag1st}-\eqn{lag2nd}, 
the ordering inconsistency occurs.
One needs to take into account the difference between $\Gamma_\nu$ and  $\gamma_\nu$.
This is because the parametric limit of the derivative of a function is not equivalent to  the derivative of the limit
function. Equation \eqn{ii7.1ft} does not apply to the systems containing a rapidly varying coordinate.
To correct the transformation rule, we continue
 the derivation of the transformation rule in \eq{ii7.d1f0} (or  \eq{ii7.for}) to include the transformation from
 $\Gamma_\nu$ to $\gamma_\nu$ in \eq{bigsmall}. This is carried out as follows
\bean
&&- g^\lambda(\vec  Z) \lbs      \pd {\Gamma_\nu(\vec Z,\epsilon)} ,{Z^\lambda}
-   \pd {\Gamma_\lambda(\vec Z,\epsilon)} ,{Z^\nu}\rbs d Z^\nu
\nn
&=& - g^\lambda(\vec Z) \lbs      \pd {\gamma_\mu (\vec Z_b(\vec Z,\epsilon))} ,{Z^\lambda} \pd Z_b^\mu,{Z^\nu}
-   \pd {\gamma_\mu (\vec Z_b(\vec Z,\epsilon))} ,{Z^\nu} \pd Z_b^\mu,{Z^\lambda} \rbs dZ^\nu. 
\eean
Using the coordinate transformation rule in \eq{ii7.ltog}, one obtains the modified transform rule
for 1 form
\bea
(L_g\bar \Gamma)_\mu &=&  g^\nu\lbs \partial_\nu\gamma_\mu    -\partial_\mu\gamma_\nu \rbs
\nn&&
-  g^\nu\lbm \lbs \partial_\nu\gamma_\delta\rbs\lbs \partial_\mu g_1^\delta\rbs   
   -\lbs\partial_\mu\gamma_\delta \rbs \lbs \partial_\nu g_1^\delta\rbs    \rbm+\cdots.
\label{ii9.for}
\eea
Here, the second term seems to be formally one order smaller than the first term on the right. 
However, if the components of $dz^\nu$ are different in order, the second term  on the right hand side of \eq{ii9.for} 
has to be kept for rapidly varying components  for ordering consistency. The gyrophase is an example of rapidly varying coordinates
in considering the charged particle motion in a magnetic field. Since the transformation rule
in \eq{ii9.for} is fundamentally important, an alternative derivation is provided in Appendix A.
  In the Appendix A, it is also pointed out that  the conventional transform rule for 1 form reviewed in the previous section
is actually obtained from the formula $\Gamma_\mu dZ^\mu =\gamma_\mu dZ^\mu$, instead of 
$\Gamma_\mu dZ^\mu =\gamma_\mu dz^\mu$. This is the consequence of the   non-consistent exchange 
 between the limit ($\epsilon\to 0$) and derivative.

The phase space Lagrangian for a charged particle motion in the electromagnetic field is given in references  \cite{littlejohn83,cary09}
\bean
^d\gamma&=&\lbs   \frac{e}{mc} A_\mu+ v_\mu\rbs dz^{\mu} + \lbs \frac {v^2}2 +  \frac{e}{m} \varphi\rbs dt.
\eean
Here   again, $``^d\gamma"$ has been used in order to show the individual components   of the one form $\bar\gamma$ explicitly. In the perturbation analyses,  $^d\gamma$ is expanded in $\epsilon$ as follows
\bea
^d\gamma_0&=&  \frac{e}{mc} A_\mu(z) dz^\mu, 
\label{iib7.gamma0}
\\
^d\gamma_1&=&v_\mu dz^\mu + \lbs \frac {v^2}2 +  \frac{e}{m}\varphi\rbs dt.
\label{iib7.gamma1}
\eea

In the zeroth order, one has 
\bea
^d\Gamma^{(0)} &=& dS^{(0)}  +  \frac{e}{mc}A_\mu(Z) \pd z^\mu,{Z^\nu} dZ^\nu
\nn
&=& dS^{(0)}  +  \frac{e}{mc} A_\mu(Z)  dZ^\mu  -    \frac{e}{mc}A_\mu(Z)  dg_1^\mu.
\label{lag0th1}
\eea
Here,   $dS =\partial_\mu S dZ^\mu$,  the last term on the right is kept, since $dg_1^\mu$ can be of order $1/\epsilon$ as can be proved {\it a posteriori}. 
 Letting $S_0 =  A_\mu  g_1^\mu$, one has $dS^{(0)}    -\epsilon  A_\mu  dg_1^\mu =   \epsilon  g_1^\mu d A_\mu$.
Noting that $d A_\mu$ is of order unity, one obtains
\bea
^d\Gamma^{(0)}&=&   \frac{e}{mc} \A(\X) \bcdot d \X.
\label{iib8.g0}
\eea
This derivation is in fact similar to  the direct derivation of Eq. (3.41) in reference \cite{cary09}. 

The first order Lagrangian in \eq{lag1st} becomes
\bea
^d\Gamma^{(1)} &=& dS^{(1)}  -L^{conv}_1 ~{^d}\gamma^{(0)}  + {^d}\gamma^{(1)} -  v_\mu dg_1^\mu
- g_1^\lambda  \pd \gamma^{(0)}_{\mu},{Z^\lambda} dg_1^\mu.
\label{lag1st1}
\eea
Here, $L^{conv}_1$ is the conventional operator given in \eq{ii7.for}. The last two terms 
 are the additional terms as compared to the conventional result as reviewed in \eq{lag1st}.
The fourth term on the right derives from the correction of the term $v_\mu dz^\mu$, similar to the last 
term on the right hand side of \eq{lag0th1}. The last term comes from the correction to the 1-form transformation
rule as shown in \eq{ii9.for}.

As shown in the appendix
of reference \cite{brizard98}, the conventional contribution can be reduced as fowllows
\bea
L^{conv}_1~ {^d}\gamma_0&=& -   \frac{e}{mc} g_1^i \lbs \pd\A_j,{X^i} - \pd\A_i,{X^j}   \rbs dX^j 
\nn
&=& -  \frac{e}{mc} \g_1^\X \btimes \B \bcdot d\X,
\label{iib7.lndg0}
\eea
where $\B=\bnabla\btimes \A$ has been used. Thus, the one form in \eq{lag1st1} becomes
\bea
^d\Gamma^{(1)} &=& dS^{(1)}+ \lbm (u\b+\v_\bot)-  \frac{e}{mc}\B\btimes\g_1^\X\rbm \bcdot d\X -\v\bcdot d \g^\X_1
 -  \frac{e}{mc}\g_1^\X  \bcdot\bnabla  \A \bcdot d \g^\X_1
 \nn&&
 - \lbs  \frac {u^2}2+\mu B +   \frac{e}{m}\varphi\rbs  dt.
 \label{iib8.gamma1}
\eea

Similar to the 
 direct reduction procedure in  reference  \cite{cary09}, noting further that
\bean
\frac 12 d \lbs{\g^\X_1}\bcdot\bnabla\A\bcdot{\g^\X_1}  \rbs 
&=& \frac12\lbs d  {\g^\X_1}\bcdot\bnabla\A\bcdot{\g^\X_1}  +{\g^\X_1}\bcdot\bnabla\A\bcdot d{\g^\X_1}  \rbs 
\nn&&
+  {\g^\X_1}\bcdot \lbs d \bnabla\A\rbs \bcdot{\g^\X_1},
\eean
one obtains
\bean
\g^\X_1\bcdot\bnabla\A\bcdot d{\g^\X_1} &=& -
 \frac12\lbs d{\g^\X_1}\bcdot\bnabla\A\bcdot{\g^\X_1} -{\g^\X_1}\bcdot\bnabla\A\bcdot d{\g^\X_1} \rbs
 \nn&&
 +\frac12\lbs d{\g^\X_1}\bcdot\bnabla\A\bcdot{\g^\X_1} +{\g^\X_1}\bcdot\bnabla\A\bcdot d{\g^\X_1} \rbs
 \nn
 &=&-\frac12\lbs d{\g^\X_1}\bcdot\bnabla\A\bcdot{\g^\X_1} -{\g^\X_1}\bcdot\bnabla\A\bcdot d{\g^\X_1} \rbs
 +\frac12 d \lbs{\g^\X_1}\bcdot\bnabla\A\bcdot{\g^\X_1}  \rbs
 \nn&&
  - {\g^\X_1}\bcdot \lbs d \bnabla\A\rbs \bcdot{\g^\X_1}.
 \eean
 Excluding the exact derivative and $O(\epsilon)$ terms, one has
 \bea
{\g^\X_1}\bcdot\bnabla\A\bcdot d{\g^\X_1} &=& - \frac12{\g^\X_1} \bcdot
\lbs d{\g^\X_1}\bcdot\bnabla\A -{\g^\X_1}\bcdot\bnabla\A \rbs
 \nn
 &=& \frac12{\g^\X_1}\btimes d{\g^\X_1}\bcdot\B + O(\epsilon).
 \label{iib5.vector}
\eea
Here, it has been noted that $ d \g^\X_1 \btimes \bnabla\btimes\A =\bnabla\A\bcdot d\g^\X_1 - d\g^\X_1\bcdot\bnabla\A$.
Therefore, one form in \eq{iib8.gamma1}  is
further reduced to
\bea
^d\Gamma^{(1)} &=& dS^{(1)} +\lbm (u\b+\v_\bot)-  \frac{e}{mc}\B\btimes\g_1^\X\rbm \bcdot d\X -\v\bcdot d \g^\X_1
 - \frac12  \frac{e}{mc}{\g^\X_1}\btimes d{\g^\X_1}\bcdot\B
 \nn&&
 - \lbs  \frac {u^2}2+\mu B +  \frac{e}{m} \varphi\rbs  dt.
 \label{iib8.gamma2}
\eea

Again similar to the 
 direct reduction procedure in  reference  \cite{cary09}, one can see that, to reduce $^d\Gamma_1$ 
one can choose
\bea
\g_1^\X&=& -{\bfg \rho},
\label{iib8.g1}
\eea
  where $-{\bfg \rho} = \v\btimes \b/\Omega$.
In this case, the term $\v_\bot \bcdot d\X$ term is cancelled,  the term $\v\bcdot d \g^\X_1$ becomes
of order $\epsilon$, and the term $- \frac12  \frac{e}{mc}{\g^\X_1}\btimes d{\g^\X_1}\bcdot\B$ is reduced to
$ (e/mc)\mu d\zeta$. Combing the contributions from $^d\Gamma^{(0)}$ and  $^d\Gamma^{(1)}$, one finally obtains from \eq{iib8.gamma2}
\bea
 ^d\Gamma = \lbs    \frac{e}{mc}\A +  u\b\rbs \bcdot d\X
+ \frac{mc}e\mu d\zeta - \lbs  \frac {u^2}2+\mu B +   \frac{e}{m}\varphi\rbs  dt.
\label{iib8.lf}
\eea
  This is the result obtained with the modified transform rule in \eq{ii9.for}.

Equation \eqn{iib8.lf} agrees with the result using the direct approach in \eq{iib8.l4} \cite{cary09},
 This is different from the conventional results in references \cite{littlejohn83,brizard98} ,
 as cited in Eqs. \eqn{zero} and \eqn{first},
in which the same result   (i.e., the term $  (mc/e)\mu d\zeta$) is only obtained in the
second order, instead of the first order.
  As indicated in the ordering analyses in \eq{p3},
the conventional result in \eq{first} with  $u\b \bcdot d\X$ being kept  but the term  $  (mc/e)\mu d\zeta$  dropped
is ordering inconsistent.

Equation \eq{iib8.lf} is obtained through the modified transformation rule for one form
in \eq{ii9.for}. Two key factors are taken into account in deriving \eq{ii9.for}.
 First, the ordering difference between  the temporal variation of gyrophase and that of the other 
phase space coordinates needs to be taken into account. Second, 
it is also important to note that the limit and derivative cannot be commuted in general.
One can expect that not only the transformation rule for one form is changed, 
but also any higher forms involving 
$dz^\mu$ (for example 2 or 3 forms, etc..) are affected as well. Nevertheless,
it is noteworthy to note that 
 since the main change for guiding
center motion theory lies in the transformation rule for one form, i.e., the Lagrangian, and one
usually does not need a backward transformation for Lagrangian. The usual transformation
rule for zero form remains basically valid. This leads the forward and backward guiding center
coordinate transformation in the conventional Lie transform formulation in \eq{zzz0} remain applicable.
This indicates that the backward coordinate transformation can still be easily obtained by
the inversion of the near identity exponential operator $e^{-\epsilon L_g}$ from the forward transformation.    
Therefore, the modified framework for the case with a fast varying coordinate
remains convenient for practical applications.

\section{Conclusions and discussion}

In this paper, we show that the conventional  Lie transform perturbation theory
for the guiding center motion of charged particles in a magnetic field needs to be
modified for ordering inconsistency. The reasons are two folds. First,
the components of $dz^\mu$ can be different in order. 
In the case of the guiding center motion of charged particles in a magnetic field,
 the temporal variation of gyrophase is much faster than other 
phase space coordinates. This is actually noted in the non-Lie-transform formulation
in reference \cite{cary09}. The other is related to the basic calculus rule. The parametric
limit of the derivative of a function is not equivalent to  the derivative of the limit
function. This leads to the change of the near identity transformation rule.

With this ordering correction made, it is shown that the Lie transform approach
can achieve the same phase space Lagrangian of guiding center motion in the first order
as obtained by the direct approach in reference \cite{cary09}. Without the correction, 
the same  Lagrangian can only be obtained by the expansion in 
the second order and beyond  \cite{littlejohn83,brizard98}.   
This is a mathematical problem. Its results were confirmed in multiple ways. The recovery of the results
with direct approach in \eq{iib8.l4} is a direct justification.  
The ordering analyses in Eq. (3) justify the current result in \eq{iib8.lf}, instead of  the conventional one in \eq{first}. Also, the key result in the paper,  i.e., the transformation rule for one form in Eq.(39), is double confirmed by an alternative derivation in Appendix A.

Let's discuss this further. The conventional Lie transform  theory \cite{littlejohn83,brizard98} 
actually belongs to the regular perturbation theory.
Because $d\X$ and $d\zeta$ are different in order, there needs to be a singular perturbation theory.   
This is similar to  the  treatment of
boundary layer problem  in the fluid theory,   the boundary layer theory applied to tearing modes
in plasma physics \cite{furth},
and the renormalization process  in dealing with  the divergent issue in the quantum field theory. 
In many fields of physics, people have experienced such type of   perturbation theory evolution.
 When the regular perturbation theory 
 was found to be incorrect, the singular perturbation theory is developed with some kind of ``renormalization".
 In this regard,   J. R. Cary and   A. J. Brizard  made an important contribution in Ref. \cite{cary09}.

The importance of current work lies in the modification of the transformation rule for one form or the 
Lagrangian for the system with a fast-varying coordinate.
In mathematics, differential forms give a unified description for defining integrands over curves, surfaces, solids, and higher-dimensional manifolds \cite{iib.arnold89,form}. It has many applications, especially in physics, geometry, and topology. This certainly includes plasma physics due to pioneer contributions by 
R. G. Littlejohn, J. R. Cary,  A. J. Brizard, et al. \cite{cary83}. For example, Hamilton's principle of least action is directly related to one form. The physics system often needs perturbation analyses. The Lie transform formalism for near identity transformation in the phase space provides a unique and powerful tool for analyzing the Lagrangian system. In plasma physics, it has been used to study 
the charged particle motion in a magnetic field, the nonlinear gyrokinetics,  the magnetic field flow, et al.
 (see for instance the review articles in Refs. \cite{cary83,brizard07,cary09}).
 The Lie transform perturbation theory deals with  the variations of various forms, i.e., the integrands,  with the change of the variable differentials taken into account.  
 The change of the one-form transform rule pointed out in this paper is critically important
to the Lie transform framework. 
To justify the need for this modification on a solid basis,
 the charged particle motion   in a magnetic field  is used as an example  for demonstration
 since it can be compared with the results derived directly and  verified by the obvious ordering analyses. 
The correction to the near identity transformation rule  affects
not only the theory of the guiding center motion of charged particles in a magnetic field, 
but also generally to the systems with a rapidly varying coordinate.
In plasma physics, for example, one can see that the earlier derivation of the nonlinear gyrokinetic equation 
needs to be repaired since the last term in the newly derived transform rule in \eq{ii9.for} has not been taken into consideration. This also affects the applications in the classical mechanics
and even the fundamental mathematical formulation of Lie transformation as well.  
The principle for the modification of one form pointed out in this paper affects the transform rules
for other forms, for example the two form related to the exterior derivative.
Again, two factors are needed to consider for other forms in the perturbation analyses.
 First,  one cannot just treat the integrands, i.e., various forms, 
 order by order,  but ignore the ordering difference between the variable differentials in the integration.  Second, one cannot simply commute the limit and derivative.  
This requires a
 systematical reformulation of  Lie transform  for higher forms and will
be  addressed in the future work. 
Nevertheless, for studying the Lagrangian system, 
the treatment of one form given in this work is usually sufficient.
These discussion indicate that the current work has a severe  impact 
on the Lie perturbation theory. It basically limits the applicability of the conventional Lie transform 
theory only to the system without a fast-varying coordinate. Note that  one of the most important
applications of Lie transform is to treat the system with a fast-varying coordinate 
to obtain the averaged effects over the fast-varying coordinate.
These indicate that the results in this paper are important. 

This research is supported by Department of Energy Grants DE-FG02-04ER54742.

\appendix

\section{Alternative derivation}

\label{app}

In this Appendix, we provide an alternative derivation of  the transformation rule
in \eq{ii9.for}. Note that
\bea
\Gamma_\mu dZ^\mu =\gamma_\mu dz^\mu.
\label{a1}
\eea
Using the transformation  rule for scale function in \eq{ii7.sde}, one obtains
\bea
\pd  ,\epsilon  \lbs  \Gamma_\mu dZ^\mu\rbs& =&\pd  ,\epsilon  \lbs \gamma_\mu dz^\mu\rbs
\nn
&=& \gamma_\mu d\pd z^\mu,\epsilon +\pd \gamma_\mu,\epsilon  d z^\mu
\nn
&=&- d\gamma_\mu \pd z^\mu,\epsilon +\pd \gamma_\mu,\epsilon \pd z^\mu,{Z^\nu} d Z^\nu +d\lbs \gamma_\mu \pd z^\mu,\epsilon\rbs
\nn
&\to&-\pd \gamma_\mu,{Z^\nu}  \pd z^\mu,\epsilon    d Z^\nu    +\pd \gamma_\mu,\epsilon \pd z^\mu,{Z^\nu} d Z^\nu.
\label{a2}
\eea
Here, the total derivative has been dropped, since we consider the variational principle. 
Using Eqs. \eqn{ii7.bde} and \eqn{ii7.dgd}, one has
\bea
\pd  ,\epsilon  \lbs  \Gamma_\mu dZ^\mu\rbs&=&\pd \gamma_\mu,{Z^\nu}  \lbs g^\lambda \pd z^\mu,{Z^\lambda}  \rbs     d Z^\nu  
  -g^\lambda \pd \gamma_\mu,{Z^\lambda} \pd z^\mu,{Z^\nu} d Z^\nu
\eea
Carrying out the expansion for $\partial {z^\nu}/\partial Z^\nu$ to sufficient order, the transformation rule for one form
in \eq{ii9.for} is recovered.

Similar to the derivation of \eq{a2}, it can be shown that the conventional result
\bean
\pd  ,\epsilon  \lbs  \Gamma_\mu dZ^\mu\rbs&=& g^\mu    \lbs  \pd \gamma_\nu,{Z^\mu} -\pd \gamma_\mu,{Z^\nu}     \rbs d Z^\nu
\eean
is actually obtained from the formula $\Gamma_\mu dZ^\mu =\gamma_\mu dZ^\mu$, instead of 
$\Gamma_\mu dZ^\mu =\gamma_\mu dz^\mu$ in \eq{a1}. This is the consequence of the   non-consistent exchange 
 between the limit ($\epsilon\to 0$) and derivative.

\newpage

\end{document}